\documentclass[a4paper,11pt,aps,pra,amsmath,amssymb,amsfonts,superscriptaddress,notitlepage,reprint]{revtex4-1}

\usepackage{graphicx} 
\usepackage{hyperref}
\usepackage{lipsum}
\usepackage{color} 
\usepackage{lmodern}
\usepackage{textcomp}
\usepackage[T1]{fontenc} 

\newcommand*{\ket}[1]{|{#1}\rangle}
\newcommand*{\bra}[1]{\langle{#1}|}

\def\be{\begin{equation}}
\def\ee{\end{equation}}
\def\bes{\begin{equation*}}
\def\ees{\end{equation*}}

\newcommand{\Pin}{P_{\rm in}}

\newcommand{\omge}{\omega_{ge}}

\newcommand{\nge}{n_{r}^{\rm ge}}
\newcommand{\nef}{n_{r}^{\rm ef}}
\newcommand{\nth}{n_{r}}
\newcommand{\Trad}{T_{\rm r}}
\newcommand{\Tmx}{T_{\rm MX}}

\newcommand{\deph}{\gamma _{\phi}}

\newcommand{\Grat}{\Gamma/|\alpha|}  

\newcommand{\Gnr}{\Gamma_{\rm nr}}

\newcommand{\NET}{\mathbb{NET}}
\newcommand{\NETP}{\mathbb{NETP}}
\newcommand{\NEP}{\mathbb{NEP}}

\newcommand{\wwin}{w_{\rm win}}
\newcommand{\Gdut}{\Gamma_{\rm DUT}}
\newcommand{\reres}{\mbox{Re}[r(0)]}
\newcommand{\Lorfun}{f_{L}}
\newcommand{\Liou}{\mathcal{L}}
\newcommand{\cohamp}{\beta}

\begin{document}

\title{Primary thermometry of propagating microwaves in the quantum regime}
\author{Marco Scigliuzzo}
\email{scmarco@chalmers.se}
\author{Andreas Bengtsson}
\affiliation{Department of Microtechnology and Nanoscience, Chalmers University of Technology, 412 96 Gothenburg, Sweden}
\author{Jean-Claude Besse}
\author{Andreas Wallraff}
\affiliation{Department of Physics, ETH Zurich, CH-8093 Zurich, Switzerland}
\author{Per Delsing}
\affiliation{Department of Microtechnology and Nanoscience, Chalmers University of Technology, 412 96 Gothenburg, Sweden}
\author{Simone Gasparinetti}
\email{simoneg@chalmers.se}
\affiliation{Department of Microtechnology and Nanoscience, Chalmers University of Technology, 412 96 Gothenburg, Sweden}
\affiliation{Department of Physics, ETH Zurich, CH-8093 Zurich, Switzerland}

\date{\today}

\begin{abstract}

The ability to control and measure the temperature of propagating microwave modes down to very low temperatures is indispensable for quantum information processing, and may open opportunities for studies of heat transport at the nanoscale, also in the quantum regime.
Here we propose and experimentally demonstrate primary thermometry of propagating microwaves using a transmon-type superconducting circuit. Our device operates continuously, with a sensitivity down to $4\times 10^{-4}$ photons/$\sqrt{\mbox{Hz}}$ and a bandwidth of $40~\rm{MHz}$. We measure the thermal occupation of the modes of a highly attenuated coaxial cable
 in a range of 0.001 to 0.4 thermal photons, corresponding to a temperature range from 35 mK to 210 mK at a frequency around 5 GHz.
To increase the radiation temperature in a controlled fashion, we either inject calibrated, wideband digital noise, or heat the device and its environment. 
This thermometry scheme can find applications in benchmarking and characterization of cryogenic microwave setups, temperature measurements in hybrid quantum systems, and quantum thermodynamics.

\end{abstract}

\maketitle

\begin{figure}%
\includegraphics[width=\linewidth]{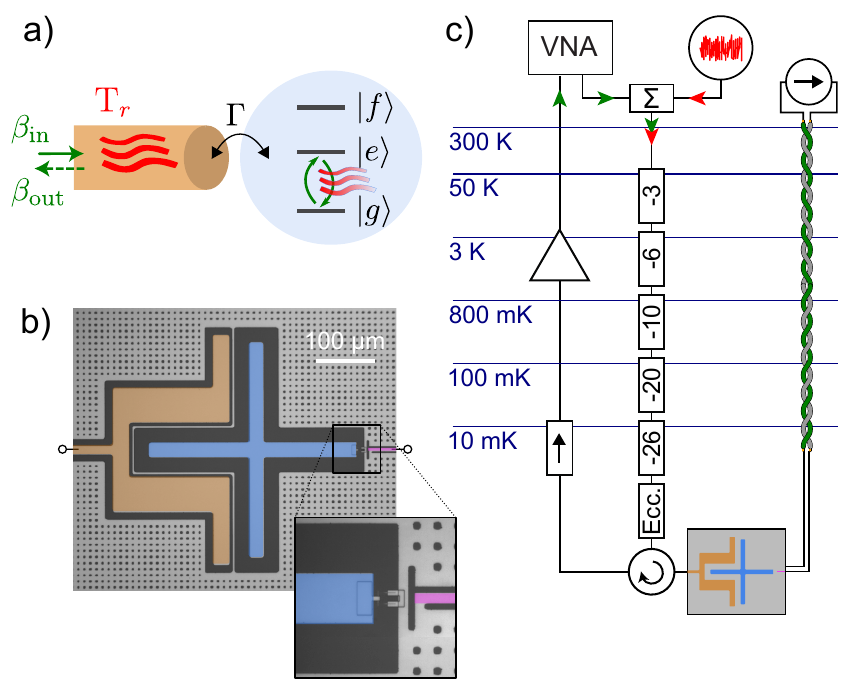}
\caption{Primary thermometry of propagating microwaves.
(a) Concept. A ladder-type emitter is strongly coupled (with rate $\Gamma$) to the end of a waveguide. We monitor the scattering of a weak coherent tone, $\cohamp_{\rm in}$ (green straight arrow), resonant with the ground-to-first-excited-state transition of the emitter. Thermal photons in the waveguide at temperature $T_r$ (wavy red lines) disturb the coherent oscillations of the emitter, leading to a decrease in the amplitude of the coherently scattered signal, $\cohamp_{\rm out}$ (green dashed arrow).
(b) False-color micrograph of the superconducting circuit realizing the setting in (a): a transmon consisting of a superconducting island (blue) shunted by a Superconducting QUantum Interference Device (SQUID) is capacitively coupled to a coplanar waveguide (orange) and inductively coupled to a flux line (purple). The inset shows a close-up of the SQUID and the flux line.
(c) Simplified scheme of the experimental setup (see text for details and Appendix~\ref{app:Exp} for the full wiring diagram).
}
\label{fig:concept}%
\end{figure}

\section{Introduction}

Propagating modes of microwave waveguides play a key role in quantum information processing with superconducting circuits, connecting the quantum processor with the classical electronics controlling it \cite{Krinner2019}.
Thermalization of these modes to the lowest achievable temperature is essential to limit unwanted decoherence of the quantum bits (qubits). 
At the same time, superconducting circuits are considered as a viable platform for thermodynamics experiments in the quantum regime \cite{Masuyama2018,Cottet2017,Ronzani2018,Naghiloo2018,Senior2019,Binder2018,Uzdin2015}. From that perspective, microwave waveguides, hosting a continuum of modes with an Ohmic spectral density, are natural candidates for realizing heat baths. Indeed,
their use as (ideally) zero-temperature reservoirs has been pioneered in quantum state preparation and stabilization protocols based on dissipation engineering \cite{Murch2012a,Kapit2017}.

For these and other applications, a fast and accurate estimation of the thermal occupation of propagating microwave modes is highly desirable.  An established method to obtain this estimate in steady state \cite{Bertet2005,Rigetti2012,Sears2012,Yan2016a,Yan2018,Yeh2017,Wang2019,Wang2019o} relies on Ramsey measurements of the dephasing time of a qubit coupled to a cavity in the dispersive limit of circuit quantum electrodynamics. In this setting, the qubit dephasing rate, $\Gamma_\phi$, depends linearly on the average thermal occupation of the cavity, due to the statistics of thermal photon shot noise in the quantum regime \cite{Goetz2017a,Yan2018}. 
Based on an averaged measurement of a pulse sequence with a repetition time of the order of $1/\Gamma_\phi$, this method lacks in temporal resolution.
But the ability to track temperature changes in real time is required to explore thermal dynamics under nonequilibrium conditions and to access thermodynamic quantities such as thermal relaxation rates, heat capacity, and temperature fluctuations in small systems \cite{Gasparinetti2015,Karimi2020a}.
In the context of thermodynamics, quantum-limited radiative heating and cooling mediated by microwave photons have been demonstrated \cite{Timofeev2009,Partanen2016,Wang2019o,Xu2020}. In a series of recent experiments \cite{Tan2017a,Ronzani2018,Senior2019}, the thermal photon occupation of microwave resonators was controlled and measured with the help of embedded mesoscopic metallic resistors acting as heat reservoirs for the confined microwave modes.
Finally, efficient detectors of single propagating microwave photons have been recently demonstrated \cite{Kono2018,Besse2018,Opremcak2018,Besse2020,Lescanne2019a}, albeit with a limited bandwidth and non-continuous operation.

Here we present the concept and experimental realization of a radiation field thermometer for propagating fields, based on the coherent scattering of a quantum emitter strongly coupled to the end of a waveguide \cite{Hoi2013b}.
At vanishingly small power, zero temperature, and in the absence of additional dephasing channels, the scattering of a coherent tone at the emitter frequency is fully coherent.
Thermal photons in the waveguide reduce the coherence of the process, leading to incoherent scattering and a detectable drop in reflectance. This drop is converted into thermal occupation by a simple algebraic expression involving device parameters that can be independently measured.
This implies that this thermometry scheme is primary, in the sense that it does not require to be calibrated against another thermometer \cite{Giazotto2006}.

We demonstrate this concept, otherwise general, using a transmon-type superconducting circuit \cite{Koch2007} as the emitter. By realizing a large coupling between the transmon and the microwave waveguide to be measured, we minimize the relative influence of intrinsic decoherence mechanisms on the temperature readout. At the same time, we ensure that the transmon is in thermal equilibrium with the modes of the waveguide, as opposed to the uncontrolled environment that limits its intrinsic coherence time and would otherwise determine its steady-state thermal occupation \cite{Jin2015b,Kulikov2020}.
We probe the response of the thermometer while increasing the radiation temperature in a controlled fashion. To do so, we first inject calibrated, wideband digital noise into the waveguide, a procedure that selectively targets the  modes whose temperature is measured. Alternatively, we increase the temperature of the base plate of our refrigerator, which heats the waveguide modes but also the thermometer itself and its environment.

\section{Results}

\subsection{Concept, sample, and characterization}

We implement the radiation field thermometer as a superconducting, frequency-tunable, transmon-type artificial atom \cite{Koch2007,Barends2013} coupled to the end of a coplanar microwave waveguide [scheme in Fig.~1(a) and micrograph in Fig.~1(b)]. The frequency of the transmon is tuned by the magnetic flux threading its Superconducting QUantum Interference Device (SQUID) loop, which we control by applying a dc current to an on-chip flux line. We operate the transmon at its highest fundamental frequency $\omega_{ge}/2\pi=5.332~\rm{GHz}$, commonly referred to as a ``sweet spot'', to minimize flux-noise-induced dephasing.
From single-tone spectroscopy (see below), we estimate the anharmonicity, $\alpha/2\pi=-217~\rm{MHz}$, and linewidth, $\Gamma/2\pi=38~\rm{MHz}$, the latter dominated by the coupling to the waveguide. We measure the coherent scattering coefficient at the waveguide input with a vector network analyzer (VNA) in a configuration comprising a highly attenuated input line, an Eccosorb filter, a circulator, a chain of isolators, and a cryogenic high-mobility electron transistor (HEMT) amplifier [Fig.~1(c); see Appendix \ref{app:Exp} for a full wiring diagram].

\begin{figure}%
\includegraphics[width=\linewidth]{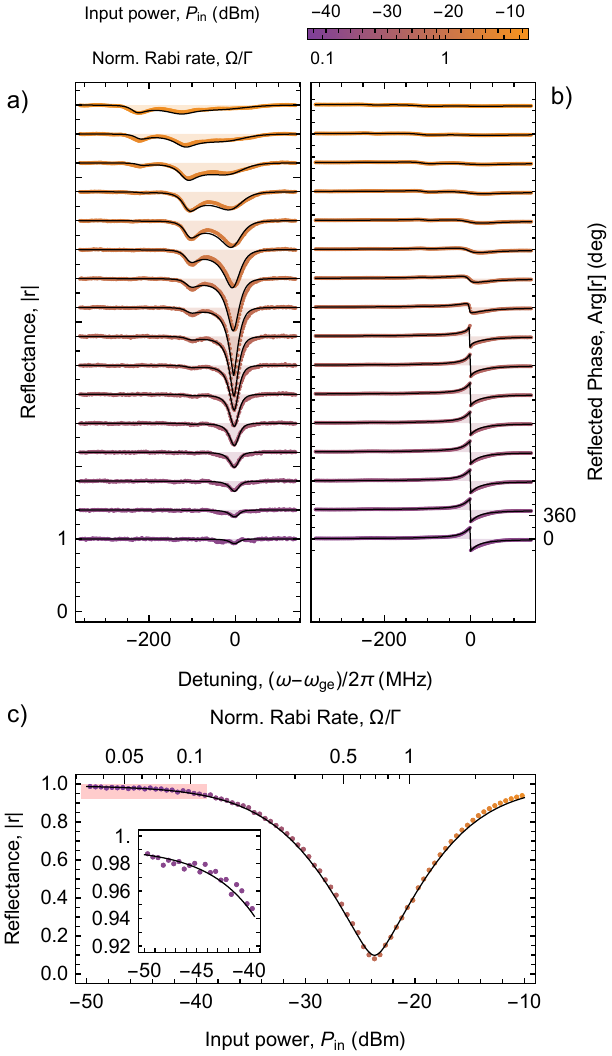}
\caption{Single-tone power spectroscopy. (a) Amplitude and (b) phase of the reflection coefficient versus drive detuning at selected input powers, $\Pin$ in the range of -39 to -9 dBm (colored dots). The traces are vertically offset by 0.4 (450 degrees) for the amplitude (phase). Solid lines show a global fit of a model to all traces (see text for details), from which sample parameters and total line attenuation are extracted. The shaded regions indicate deviations of the reflection coefficients from unit amplitude and zero phase. (c) Reflectance at resonance, $|r(0)|$, versus input power: measured values (dots) and theory (solid line). Inset: close-up of the shaded region in the main panel.
}
\label{fig:spec}%
\end{figure}

To increase the radiation field temperature, we synthesize noise in the frequency band of interest \cite{Fink2010}.
All measurements are performed in a dilution refrigerator with base temperature below 9 mK. To control the base temperature, we apply a current to heat a resistor anchored to the base plate.
We characterize our sample by reflection spectroscopy at varying input powers, $P_{\rm in}$. To set apart the response of the transmon from the frequency-dependent transfer function of the input and output line, all spectroscopy traces are referenced to a configuration in which the transmon transitions are detuned at least 1~GHz below the frequency band of interest. We have also tested a high-power trace that fully saturates the transmon as a reference. These two procedures give equivalent results for input powers $\Pin \geq 0~\rm{dBm}$, implying, in particular, that our amplification chain is negligibly compressed at this power level. The high-power subtraction scheme can be conveniently used if a fixed-frequency transmon is used as the thermometer.
With this calibration in place, we expect the most important source of inaccuracy to be related to imperfections in the circulator used to route the signals between the sample and our input and output lines. This issue is discussed in more detail in Appendix~\ref{app:circ}.

At low values of $P_{\rm in}$, the transmon is only weakly excited. Then coherent scattering dominates and the measured response is similar to that of a harmonic oscillator over-coupled to a transmission line. When sweeping the drive frequency across the resonance, the reflection coefficient shows a weak response in the amplitude [Fig.~2(a), bottom traces], and a full (180-degree) phase shift [Fig.~2(b)].
Increasing $P_{\rm in}$ increases the fraction of incoherent scattering from the transmon, resulting in a dip in the amplitude. As soon as incoherent scattering dominates over coherent scattering, the phase swing also starts to reduce. Eventually, the reflection coefficient at the fundamental resonance approaches unity as the emitter becomes saturated. In this high-power regime, however, higher-order transitions become observable as additional amplitude dips appearing at red-detuned frequencies. In particular, we clearly resolve the two-photon transition between the ground and second-excited state, $\omega_{gf}/2$, and the three-photon transition between the ground and the third excited state, $\omega_{gh}/3$, detuned by $\alpha/2$ and $\alpha$ from the fundamental transition, respectively. We note that the visibility of these higher-order transitions is enhanced by the larger ratio between linewidth and anharmonicity, $\Gamma/|\alpha|=0.18$, compared to previous work \cite{Hoi2013b}.

We model the response of the system by a master equation and input-output theory, taking up to 4 levels of the transmon into account (for more details on the simplifying assumptions, we refer to Appendix \ref{app:Th}). 
In our analysis we extract the values of $\omega_{ge}$, $\Gamma$, $\alpha$, and $A$.
This calibration allows us to convert the input power into a drive rate $\Omega$, via the relation
\be
\Omega = 2 \sqrt{A \Gamma  \Pin /\hbar \omge} \ .
\ee
The global fit of our model to the spectroscopy data [Fig.~2(a,b), solid lines] presents an excellent agreement across a power range of four orders of magnitude. The inclusion of the fourth level is necessary to described the third-order transition $\omega_{gh}/3$, the agreement being otherwise excellent also for a three-level model.

\subsection{Thermometry concept}

We now focus our attention on the power dependence of the magnitude of the reflection coefficient at resonance, $|r(0)|$ [Fig.~2(c)]. The observed V-shaped curve arises from the coherent interference between the radiation scattered by the emitter and that reflected by the open end of the waveguide, which carry opposite phase \cite{Hoi2015,Wen2018}. Note that a dip in the magnitude is not associated with a decrease in the total reflected power, but only with the coherent part of it; in our case, we estimate (based on the coupling rates given below) that over $99.99\%$ of the power is always reflected. The measured $|r(0)|$ reaches a minimum of $\approx 0.1$ when the ratio of the drive rate over the linewidth is about $\Omega/\Gamma \approx 1/\sqrt{2}$.
For a pure two-level system, the coherent reflection at resonance would vanish at this drive rate. For a transmon, complete destructive interference (cancellation) can still be attained, but the drive parameters at which this happens are affected by the presence of the third level. In particular, full cancellation is expected at a frequency detuned by $\Gamma^2/2\alpha\approx -3~\rm{MHz}$ (to first order in the ratio $\Gamma/|\alpha|$), which explains why the cancellation exhibited by the dip in Fig.~2(c) is only partial.

In the low-power limit $\Omega\ll\Gamma$, $|r(0)|$ approaches the asymptotic value $0.988$, deviating from unity by $0.012$ [Fig.~2(c), Inset]. To explain this drop in reflectance, we consider three possible mechanisms: (i) decay to modes other than the waveguide, at the (nonradiative) rate $\Gnr$, (ii) pure dephasing, at rate $\deph$, and (iii) thermal occupation in the waveguide, $\nth$. We calculate the first-order contributions of these mechanisms to the reflection coefficient to be
\be
r(0)=  - 1 +\frac{12}{1+3i\Gamma/2|\alpha|} \nth + 4 \frac{\deph}{\Gamma} + 2\frac{\Gnr}{\Gamma}
+ \frac{4(\Omega/\Gamma)^2}{1+i\Gamma/|\alpha|}
\ .  \label{eq:r0} 
\ee
From Eq.~\eqref{eq:r0}, we notice that the impact of decoherence mechanisms is largely mitigated by our design choice of a large coupling to the waveguide. At the same time, the finite $\Grat$ ratio renormalizes the conversion factor linking reflection coefficient to thermal occupation. In the limit $\Grat \to 0$, the thermal occupation is related to the measured quantity, $r$, by an algebraic expression involving only numerical constants. Equation~\eqref{eq:r0} also gives the first correction to the reflection coefficient due to finite input power (saturation), which is of second order in the ratio $\Omega/\Gamma$.

Based on measurements of transmon qubits fabricated on the same wafer \cite{Bengtsson2019}, similar to those reported in Ref.~\cite{Burnett2019}, with identical geometry but no direct coupling to a waveguide, we estimate 
$\Gnr/2\pi<4~\rm{kHz}$ ($T_1>40~\mu\rm{s}$) and $\deph/2\pi<2~\rm{kHz}$ ($T_2^*>60~\mu\rm{s}$).
(These estimated rates are $10^{4}$ times smaller than the decay rate into the waveguide, therefore a direct measurement of them \cite{Lu2019f} is challenging.)
According to \eqref{eq:r0}, we thus expect the combined pure dephasing and intrinsic decay mechanisms to cause a drop $\delta |r(0)|<0.0004$, much smaller than the observed one. We thus conclude that the observed drop is most likely due to thermal occupation in the waveguide, $\nth=(1.0 \pm 0.3) \times 10^{-3}$, corresponding to an effective photon temperature $\Trad=37~\rm{mK}$. To gain more confidence in the validity of this analysis, we have performed the measurements with added noise reported below.
For comparison, we estimate the expected temperature of the waveguide in the ideal case based on the attenuation configuration of the line \cite{Krinner2019}, assuming perfect thermalization of the components at their stages (Appendix~\ref{app:Exp}), and find $\Trad^{\rm est}=27~\rm{mK}$ (corresponding to $\nth^{\rm est}=8\times 10^{-5}$).
The measured radiation temperature is thus higher than what would be expected for an ideal chain of attenuators.

\subsection{Thermometry with added noise}

\begin{figure}%
\includegraphics[width=\linewidth]{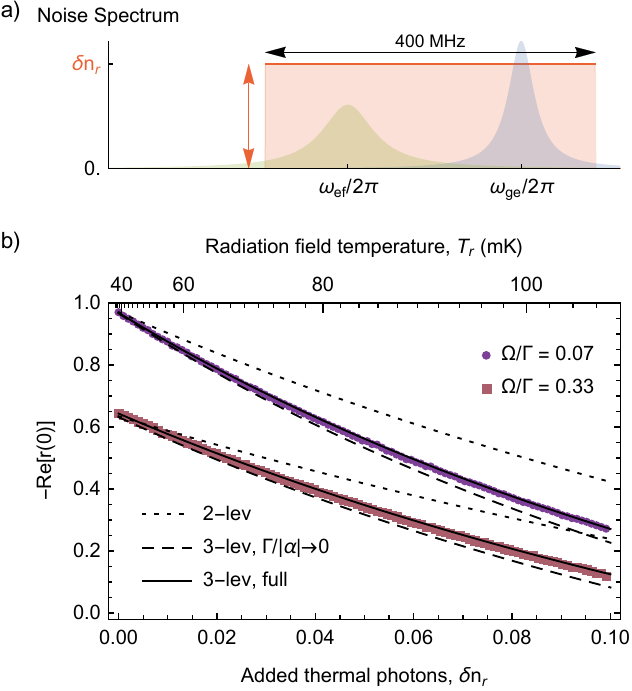}
\caption{Radiation field thermometry with artificial white noise.
(a) Sketch of the power spectral density of the applied noise (solid red line), with the relevant transmon transitions indicated on the frequency axis. The noise is white over a frequency band of 400 MHz and has a calibrated spectral density $\delta\nth$. Blue and green shades indicate the spectral densities of the transmon transitions, with a relative scaling reflecting their participation to the thermometry signal.
(b) Measured real part of the reflection coefficient at resonance, $-\reres$, at two selected drive strengths, $\Omega/\Gamma=0.07\mbox{ and } 0.33$, versus calibrated noise power expressed in added thermal photons, $\delta n_r$. The data (symbols) is directly compared to increasingly refined models, without free parameters: two-level theory (short dashes), three-level theory assuming a small linewidth-to-anharmonicity ratio, $\Gamma/|\alpha|\to 0$ (long dashes), and three-level theory taking the measured ratio $\Gamma/|\alpha|=0.18$ into account (solid line).
}
\label{fig:noiseampsweep}%
\end{figure}

To confirm our ability to measure the temperature of the radiation field in the waveguide, we increase it in a controlled fashion by applying white noise of adjustable rms amplitude over a 400 MHz bandwidth comprising the first two transitions of the transmon [Fig.~\ref{fig:noiseampsweep}(a)]. We take into account the finite noise bandwidth introducing a windowing coefficient $\wwin=0.89$, determined by the overlap between the noise spectral density and the linewidths of the $g$-$e$ and $e$-$f$ transitions of the transmon, with a relative weight given by the participation of these transitions to the expected response in reflection [see Appendices \ref{app:Th} and \ref{app:Analys}]. 
Using the calibrated attenuation of the line from the spectroscopy data, as well as a room-temperature calibration of the noise power compared to the power of the coherent tone use for the spectroscopy, we determine the added thermal noise in photons, $\delta\nth$, as a linear function of the noise power.

We study the response of the real part of the reflection coefficient at resonance, $\reres$ to the added thermal noise at low ($\Omega/\Gamma=0.07$) and intermediate ($\Omega/\Gamma=0.33$) drive powers 
[Fig.~\ref{fig:noiseampsweep}(b)]. For both drive powers, an increase in $\delta\nth$ produces a drop in $-\reres$. The slope of the curve, $\partial \reres/\partial \delta\nth$, gives the responsivity of the thermometer as a function of the photon occupation, and is larger for the low-power trace. We discuss the tradeoff between power and responsitivity in Section \ref{ssec:Sens} (see also Fig.~\ref{fig:SensResp}).

Our model for the drop in $-\reres$, based on the parameters extracted from spectroscopy, agrees well with the data at both powers [Fig.~\ref{fig:noiseampsweep}(b), solid lines]. In particular, a best-fit of the linear slope of the low-power trace for $\delta\nth\ll1$ agrees with the predicted one within one percent. This confirms the validity of the calibration scheme that we used, based on acquiring a reference trace with the transmon detuned: a systematic error in the calibration would produce a different slope. Conversely, if this scheme is not available or cannot be trusted, data sets such as the ones in Fig.~3 can itself serve as a calibration.

For comparison, we also show the model for a two-level system (short dashes), and for a three-level transmon in the limit $\Gamma/|\alpha|\to 0$ (long dashes). The comparison against a two-level system highlights the fact that the second excited state $\ket{f}$ contributes to the response even at the smallest measured photon numbers, for which we expect its steady-state population to be negligibly small, $P_{f}\approx 10^{-5}$. Indeed,
noise spectroscopy measurements [see Appendix~\ref{app:NoiseSpec}]
confirm that the thermometer responds to thermal photons at both the fundamental and the first-to-second-excited state transition, in agreement with the model.
Finally, our comparison highlights the necessity to consider the finite value of the $\Grat$ ratio to reproduce the observed response. Since the $\Grat$ ratio can be accurately determined from spectroscopy, however, this fact does not compromise the primary nature of the thermometer.

\subsection{Thermometry at different system temperatures}

\begin{figure}%
\includegraphics[width=\linewidth]{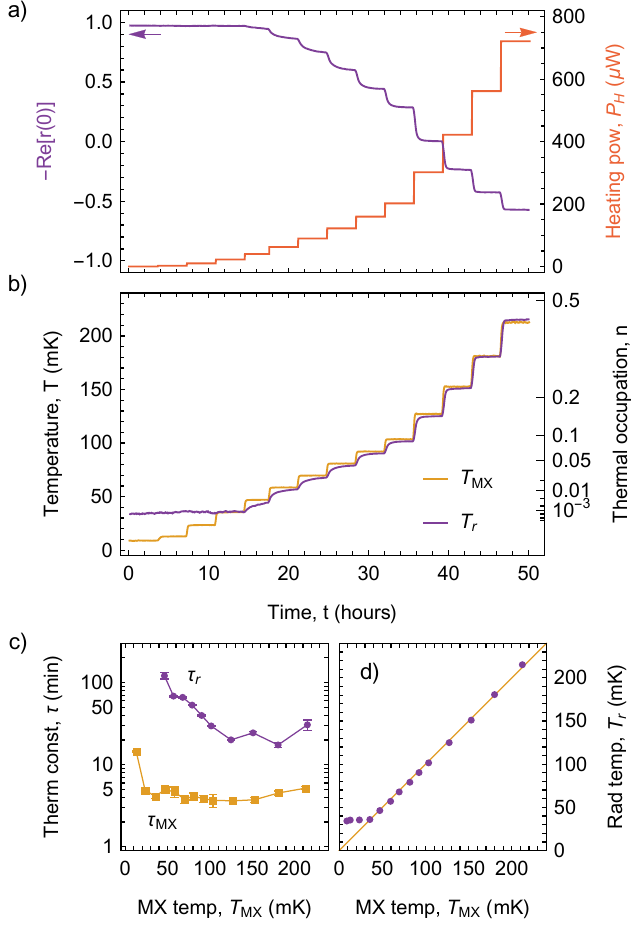}
\caption{Radiation field thermometry while varying the temperature of the base plate.
(a) Temporal profile of the heating power, $P_H$ (right axis, red) used to vary the base-plate temperature $\Tmx$, and corresponding temporal variation of the real part of the reflection coefficient (left axis,purple).
(b) Corresponding base-plate temperature, $\Tmx$ (yellow) and radiation-field temperature, $\Trad$ (purple) versus time. (c) Thermal relaxation constants $\tau_{\rm MX}$ (squares) and $\tau_r$ (circles) for $\Tmx$ and $\Trad$, respectively, extracted from the data in (b) and plotted versus the steady-state $\Tmx$. The solid lines are guides for the eye.
(d) Steady-state values of $\Trad$ versus $\Tmx$ (circles); the solid line indicates $\Trad=\Tmx$.
}
\label{fig:Tsweep}%
\end{figure}

We vary the temperature of the radiation field in another, independent way by heating the base plate of the dilution cryostat. Differently from the noise injection experiment, in this case the full system is warmed up, including the modes of the waveguide, which we expect to thermalize via the attenuators anchored to the base plate, but also the environment in which the thermometer resides, and in particular, any spurious baths to which the transmon may be coupled.

We adjust the heating power to the base plate of the cryostat, in time steps of 3 hours, and monitor the response of $-\reres$ [Fig.~\ref{fig:Tsweep}(a)]. Consistently with the previous experiment, we observe that an increase in temperature results in a reduction of $-\reres$. Compared to Fig.~\ref{fig:noiseampsweep}, here we study a larger variation in temperature, and measure a correspondingly larger variation in $-\reres$, in the range of 0.99 to $-0.6$. Notice that the relation between $-\reres$ and temperature is monotonic (and, therefore, single-valued), even when the sign of $-\reres$ changes from positive to negative.

Using the full theoretical expression for the reflection coefficient (Appendix~\ref{app:Th}), we extract the thermal occupation, convert it into radiation temperature $\Trad$, and compare it to the temperature of the base (mixing chamber) plate, $\Tmx$, measured independently by a calibrated Ruthenium oxide thermometer [Fig.~\ref{fig:Tsweep}(b)]. The measured $\Trad$ is insensitive to changes in $\Tmx$ as long as $\Tmx$ is below the saturated value of $\Trad$.
This insensitivity should not be interpreted as a limitation of the thermometer, whose ability to respond to small changes in noise photon number was demonstrated in Fig.~\ref{fig:noiseampsweep}. Instead, it can be explained by considering the exponential relation between thermal occupation and temperature in this temperature range, which causes the effective photon occupation to be dominated by the highest temperature. Then it is only 
when the two temperatures become comparable that $\Trad$ starts to follow $\Tmx$.
After a change in heating power, the adjustment of $\Tmx$ to its new steady-state value has an exponential tail with a time constant, $\tau_{\rm MX}$, between 4 and 5 minutes, which does not depend on temperature in the range considered [except for the first temperature step, for which $\tau_{\rm MX}=14$~min; Fig.~\ref{fig:Tsweep}(c)]. This time scale is related to the thermalization dynamics of the base plate as a whole.
Its temperature independence results from the quadratic (linear) increase of the cooling power of the mixing chamber (heat capacity of the base plate) as a function of temperature.

By contrast, the adjustment of the radiation temperature, $\Trad$, follows a much longer time constant, $\tau_r$, which is of the order of two hours at the lowest temperature, and decreases by an order of magnitude as temperature is increased [Fig.~\ref{fig:Tsweep}(c)]. We ascribe this long time scale to a slow thermalization of the resistive section of the base-plate attenuators, whose voltage fluctuations set the temperature of the radiation field.
At temperatures below 100 mK, the thermalization of these components proceeds via a cascade process which involves electronic, phononic, and photonic channels \cite{Giazotto2006,Yeh2017,Wang2019,Yeh2020}.
The measured temperature dependence of $\tau_r$ may indicate a crossover between electron-dominated thermalization, with a thermal conductance
proportional to temperature according to the Wiedemann-Franz law, and phonon-dominated thermalization, with a thermal conductance
proportional to the fourth power of temperature for metals such as gold and copper \cite{Giazotto2006}.

We extrapolate the steady-state temperature at each step of the thermal sweep by fitting a decaying exponential to the tail of the relaxation data, and plot the asymptotic value as a function of the steady-state temperature of the base plate [Fig.~\ref{fig:Tsweep}(d)]. The two values essentially agree over more than two orders of magnitude in thermal occupations number, from $\nth=0.001 \mbox{ to } 0.4$ (corresponding to a temperature range of 35 to 210 mK).

\begin{figure}%
\includegraphics[width=\linewidth]{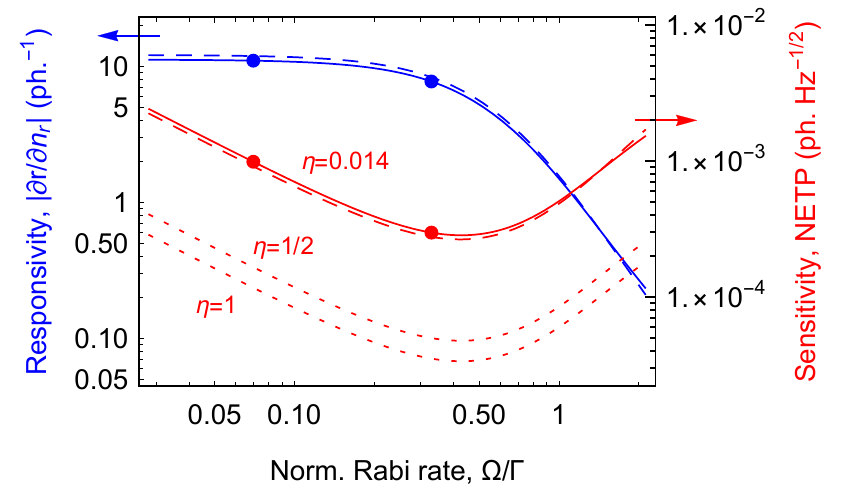}
\caption{Calculated photon number responsivity, $|\partial r/\partial n|$ (left axis), and sensitivity, $\NETP$ (in photons per square root bandwidth, right axis) versus normalized Rabi rate, $\Omega/\Gamma$, in the low-occupation limit $n\to 0$ and for the measured sample and setup parameters (solid lines). Dashed lines are calculated in the limit $\Gamma/|\alpha|\to 0$ (see text for details). The dashed lines are the projected sensitivities using the same sample and an amplification chain with efficiencies $\eta=1/2$ and $\eta=1$. Thick dots indicate the responsivity and sensitivity measured with our set-up ($\eta=0.14$) at the two driving powers used in Fig.~3.
}
\label{fig:SensResp}%
\end{figure}

\subsection{Sensitivity}\label{ssec:Sens}

We finally consider the sensitivity of this thermometry scheme. This figure is typically expressed as a noise equivalent temperature ($\NET$), in units of K/$\sqrt{\rm{Hz}}$.
However, given that the response of our detector is directly proportional to the thermal occupation $\nth$, and that the relation between $\nth$ and $\Trad$ is strongly nonlinear in the deep quantum regime, $\nth \ll 1$, we prefer to provide this figure in noise equivalent thermal photons ($\NETP$), which we calculate as
\be
\NETP=\left|\mathcal R\right|^{-1} \frac{\delta \cohamp_{\rm out}}{\cohamp_{\rm in}} \ , \label{eq:netp}
\ee
where $\mathcal R =  \frac{\partial r}{\partial \nth}$ is the responsivity of the reflection coefficient to changes in the thermal photon number, $\cohamp_{\rm in}$ is the incident coherent photon flux, and $\delta \cohamp_{\rm out}$ is the noise in the detection chain. The latter can be expressed as
$
\delta \cohamp_{\rm out} = 1/\sqrt{2\eta}
$, where the factor $1/\sqrt{2}$ is due to quantum vacuum fluctuations and $\eta$ is the quantum efficiency of the amplification chain \cite{Sete2015}. As captured by Eq.~\eqref{eq:netp}, the optimum sensitivity is obtained as a trade-off between higher responsivity, obtained at lower drive powers, and higher signal-to-noise ratio, obtained at higher drive powers [Fig.~6].
In the limit $\nth \to  0$, the optimum drive rate is found to be $\Omega/\Gamma\approx0.42$. The price to pay for an increased sensitivity is that the
responsivity depends on the normalized Rabi rate $\Omega/\Gamma$. However, since the latter can be easily calibrated [see Fig.~2], this is not a major limitation.
A large radiative linewidth $\Gamma$ is generally beneficial for the sensitivity, because it increases the photon flux that can be scattered by the transmon while keeping the ratio $\Omega/\Gamma$ constant, and therefore improves the signal-to-noise ratio.

In our experiment, we evaluate the sensitivity by measuring the standard deviation of repeated measurements at various drive rates and measurement bandwidths, extract the signal-to-noise ratio, and multiply it by the theoretical responsivity. Alternatively, we first measure the quantum efficiency of our amplification chain by comparing the power spectrum of the resonantly driven transmon at saturation with the noise level, and then use the theoretical expression.
We measure sensitivities of the order of $1\times10^{-3}~\rm{photons}/\sqrt{Hz}$ at low drive rates and $4\times 10^{-4}~\rm{photons}/\sqrt{Hz}$ close to the optimum drive rate [Fig.~\ref{fig:SensResp}, large dots]. The best sensitivity we measured is limited by our amplification chain with an added noise of 35 photons and corresponding quantum efficiency $\eta=0.014$. For an amplification chain working close to the standard quantum limit $\eta=0.5$, which has been reached and even surpassed using phase-sensitive Josephson parametric amplifiers \cite{Walter2017,Lecocq2019}, this number could be further improved by one order of magnitude [dashed line in Fig.~\ref{fig:SensResp}].
Finally, we notice that the presented device can also be operated as a narrow-band bolometer. In the current measurement setup, the corresponding noise-equivalent power ($\NEP$) is of the order of
\be
\NEP \approx \hbar \omega_{ge} \Gamma \times \NETP \approx 400~\rm{zW}/\sqrt{\rm{Hz}}\ .
\ee
For comparison, $\NEP$s in the range of a few hundred down to a few $\rm{zW}/\sqrt{\rm{Hz}}$ have been reported for nanobolometers based on metallic absorbers operating in the microwave frequency range \cite{Govenius2015a,Kokkoniemi2019}.

\section{Discussion}

In summary, we have demonstrated that a single quantum emitter strongly coupled to the end of a waveguide can be used for sensitive measurement of the thermal occupation of the waveguide, using simple continuous-wave, single-tone spectroscopy. By applying synthesized noise of controlled spectral content, we explored the response of the thermometer to a frequency-dependent thermal occupation.
Thanks to a large coupling rate of the transmon to the waveguide, we minimized the effect of the intrinsic decoherence mechanisms of the emitter on the thermometry scheme (for the numbers given in the text, their contribution to the measured thermal occupation is of the order of $1\times 10^{-4}$, one order of magnitude smaller than the lowest measured value).

The presented thermometer enables measurements of the thermal occupation of attenuated waveguides used to control quantum information processors, in a range relevant for state-of-the-art applications and with real-time capability.
Combining this device with a cryogenic microwave switch \cite{Pechal2016}, one can directly compare the performance of different filtering, attenuation, and thermalization schemes.
The reported sensitivity, together with continuous operation and a measurement bandwidth of tens of MHz, set by the radiative linewidth of the transmon, makes this thermometry scheme appealing for studies of thermodynamics at the nanoscale, both in steady state and in time domain \cite{Gasparinetti2015,Karimi2018,Karimi2020a}. The device can be used, for instance, to monitor the temperature of a radiating body such as a mesoscopic electron reservoir, to which it can be coupled via a matching circuit, or to measure the scattering of thermal microwaves from a quantum circuit acting as a quantum heat engine or refrigerator.
In future realizations, the transmon qubit may be replaced by Josephson-based artificial atoms with larger anharmonicities \cite{Yan2016a} and/or different level structures, possibly also in the ultra-strong coupling regime \cite{Forn-Diaz2017}. Non-superconducting emitters could also be considered, such as semiconductor quantum dots \cite{Stockklauser2017}.

\section{Acknowledgements}

We are grateful to F.~Marxer for his contribution to early experiments and to Yong Lu, Avgust Yurgens, Anton Frisk Kockum, and Howard Carmichael for useful discussions. We wish to express our gratitude to Lars J\"{o}nsson for making the sample holder. We are also grateful to Bertil Skoglund at RFMW for providing technical details about the cryogenic attenuators used in this work. The sample was fabricated at Myfab Chalmers. We acknowledge financial support from the Swedish Research Council, the Knut and Alice Wallenberg Foundation,
the European Union under grant agreements no. 339871 (SuperQuNet) and no. 820363 (OpenSuperQ),
the Michael Kohn Foundation, the Baugarten Foundation, and the ETH Zürich Foundation.


\appendix

\section{Experimental details}\label{app:Exp}

\subsection{Setup}
\begin{figure}%
\includegraphics[width=\linewidth]{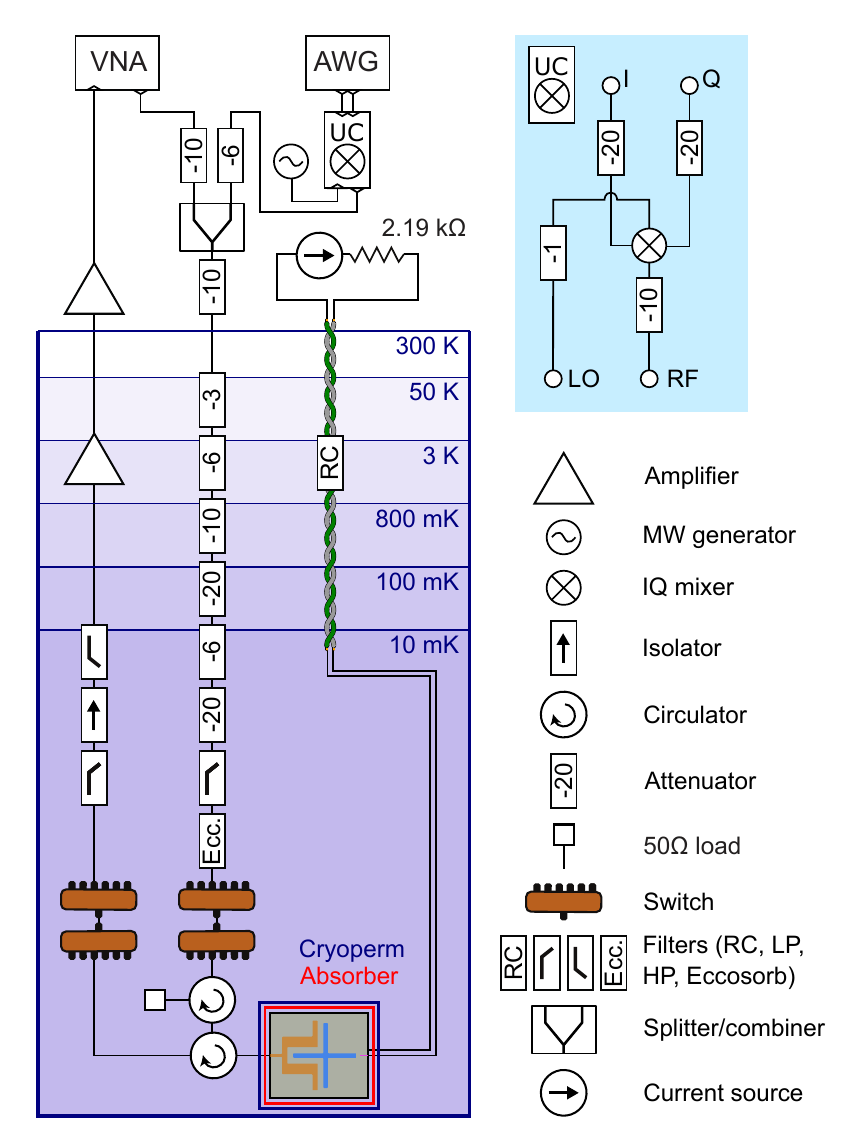}
\caption{Full wiring diagram for the experiment. VNA: vector network analyzer; AWG: arbitrary waveform generator; UC: up-conversion module; MW: microwave; LP: low pass; HP: high pass.
}
\label{fig:wiring}%
\end{figure}

The radiation field thermometer is mounted at the mixing chamber stage of a dilution refrigerator [see Fig.~\ref{fig:wiring}] and is enclosed in a non-magnetic environment, shielded by a mu-metal shield and an absorber shield.
At room temperature, digital noise is synthesized by an arbitrary-waveform generator (AWG), mixed with a local oscillator (LO) using an in-phase-quadrature (IQ) mixer, combined with a coherent probe tone from the vector network analyzer (VNA), and fed to a highly attenuated input line that includes a low-pass filter and an Eccosorb filter at base temperature. A cryogenic circulator is used to route the signal reflected from the sample to the output line, which has a cryogenic high mobility electron transistor (HEMT) amplifier mounted at the 3K stage. A high-pass filter, a low-pass filter, and four isolators are used to isolate the sample from radiation emitted by the HEMT. A dc current is fed to the flux line of the thermometer via a twisted pair of low-ohmic cables, interrupted at the 3K stage by a low-pass filter.

\subsection{Noise generation}

To generate noise with arbitrary spectral properties, we start from a string of random real numbers sampled from a normal distribution with zero mean and unit variance (white noise). We (unessentially) take the discrete Fourier transform of this signal and apply a filter centered at the intermediate frequency $\omega_{\rm IF}/2\pi=220~\rm{MHz}$, with unit amplitude and the desired spectral profile. The filter only has support at positive frequencies. We then invert the Fourier transform and send the real and imaginary part of the complex signal to the in-phase and quadrature port of the up-conversion (UC) mixer. With this procedure, the noise is up-converted to a single sideband \cite{Fink2010}.

To compensate for frequency-dependent conversion loss in the mixer, we generate white noise across the frequency band of interest, measure its spectral density at room temperature with a spectrum analyzer, and use the result to calculate a frequency-dependent transfer function for the mixer. In subsequent experiments, we apply the transfer function to the digital filter that determines the noise properties, so that the spectral density of the upconverted noise, measured at room temperature, matches the intended one.

\section{Noise spectroscopy}\label{app:NoiseSpec}

\begin{figure}%
\includegraphics[width=\linewidth]{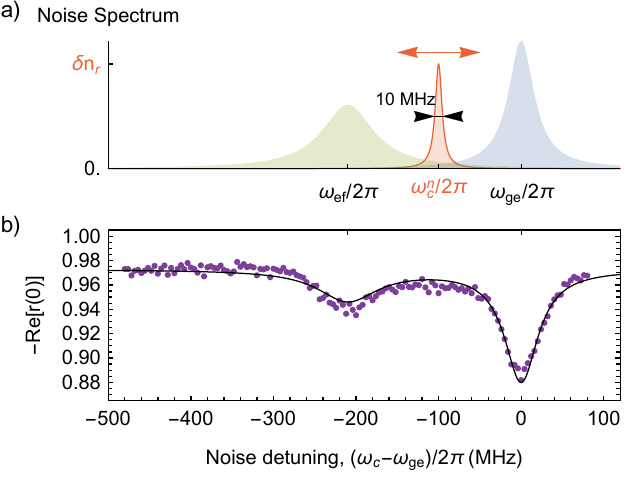}
\caption{Noise spectroscopy with Lorentzian noise.
(a) Power spectral density of the applied noise, a 10-MHz Lorentzian with fixed amplitude and varying center frequency $\omega_c/2\pi$.
(b) Real part of the reflection coefficient at resonance, $-\reres$, versus noise detuning from the fundamental transition (dots), and corresponding theory prediction (solid line) based on the same parameters as in Fig.~\ref{fig:noiseampsweep}.
}
\label{fig:Lornoisedetsweep}%
\end{figure}
To determine the relative participation of noise applied at different frequencies to the dip in reflection, we apply Lorentzian noise with varying center frequency and a linewidth of 10 MHz [Fig.~\ref{fig:Lornoisedetsweep}]. As expected, we obtain the strongest response when the applied noise is resonant with the $g$-$e$ transition. However, we observe a sizeable dip also at the $e$-$f$ resonance. Indeed, a careful theoretical analysis assigning different thermal occupations to the radiation field at the two transitions (see Appendix \ref{app:Th:refl}) reveals that both transitions participate to dephasing, with a relative strength 2:1 (in the linear-response limit, $\nth\ll1$). This ratio has a simple physical explanation, namely, that thermal photons at the $g$-$e$ transition induce both population change and dephasing, while those at the $e$-$f$ induce only dephasing on the $g$-$e$ transition (the state $\ket{e}$ is virtually unpopulated when both $\nth$ and $\Omega/\Gamma$ are small).

\section{Radiation Field Thermometer Model}\label{app:Th}

\subsection{Master equation and steady state solution}

In a frame rotating at the drive frequency $\omega_d$ and using a rotating-wave approximation, we describe the driven transmon by the Hamiltonian
\be
\hat H= - \delta \hat b^\dagger \hat b + \frac{\alpha}2 \hat b^\dagger \hat b^\dagger \hat b \hat b
+  i \frac{\Omega}{2}
(\hat{b}-\hat{b}^\dagger) \ ,
\ee
where $\hat b$ is an annihilation operator, $\delta=\omega_d-\omega_{ge}$ is the drive detuning from the fundamental resonance, $\alpha$ is the transmon anharmonicity ($\alpha<0$), and $\Omega$ is the drive strength \cite{Koch2007}.

The coupling to the waveguide induces transitions between adjacent energy eigenstates of the transmon, described by lowering and raising operators $\sigma_-^j = \ket{j}\bra{j+1}$ and $\sigma_+^j = \ket{j+1}\bra{j}$ for the $j$-th transition, respectively ($\ket{j}$ denotes the $j$-th state).
We consider frequency-dependent coupling strengths and thermal occupations, so that the waveguide modes around the $j$-th transition are coupled to the transmon with relative strength $\xi_j$, and are characterized by a thermal occupation $n_{\rm th}^{(j)}$. For our experiment, we assumed a constant density of states in the waveguide, so that $\xi_j=\sqrt{j}$ due to the dipole matrix elements of the transmon \cite{Koch2007}. We thus derive a master equation in a ``local'' Born-Markov approximation (around the frequency of each transition) \cite{Carmichael1973}. The dissipative interaction between the transmon and the waveguide is described by a Liouvillian superoperator $\Liou_\Gamma$, acting on the density matrix $\rho$ in a Hilbert space truncated to the first $N$ states:
\be
\begin{split}
\Liou_\Gamma \rho =
\Gamma \sum_{l,m=1}^{N} \xi_l \xi_m [ (1+n_{\rm th}^{(m)}) \mathcal D^-_{l,m}[\rho] 
+ n_{\rm th}^{(m)} \mathcal D^+_{l,m}[\rho]] \ ,
\end{split} \label{eq:Liou_full}
\ee
where $\Gamma$ is the decay rate of the fundamental transition. We have introduced generalized dissipators
\be
D^\pm_{l,m}[\rho]= -\frac12\left(
\rho \sigma _{\mp}^m \sigma _{\pm}^l+\sigma _{\mp}^l \sigma _{\pm}^m \rho -\sigma _{\pm}^l \rho  \sigma _{\mp}^m-\sigma _{\pm}^m \rho  \sigma _{\mp}^l\right) \ ,
\ee
which describe excitation (decay) of the $l$-th transition of the transmon, accompanied by absorption from (emission into) the waveguide of a photon with the energy of the $m$-th transition. Terms with $l \neq m$ rotate at the frequency difference between the two transitions in the interaction picture. If the linewidths of the transitions are much smaller than the difference between their frequencies, then these terms can be neglected by invoking a rotating-wave approximation \cite{Wen2018}. Here, however, due to the finite $\Grat$ ratio, these cross-terms introduce a correction that we take into account. When terms with $m \neq n$ are neglected in \eqref{eq:Liou_full}, we recover the more familiar expression
\be
\Liou_\Gamma \hat\rho(t) = 
\Gamma \sum_{j=1}^{N} \xi_j^2 \left[(1+n_{\rm th}^{(j)})\mathcal D[\sigma_-^{j},\hat\rho(t)]
+n_{\rm th}^{(j)}\mathcal D[\sigma_+^{j},\hat\rho(t)] \right] 
\ ,
\ee
with the dissipator
\be
\mathcal D[A,B] = A BA^\dagger - \frac12\left(A^\dagger A B- B A^\dagger A\right) \ .
\ee
We model pure dephasing at rate $\deph$ by adding the following Liouvillian to the master equation:
\be
\Liou_{\deph} \hat\rho(t) = \deph  \sum _{j=1}^{N} \mathcal D[(\ket{j}\bra{j},\hat\rho(t)] \ .
  \label{eq:lioudeph}
\ee
Finally, we model nonradiative decay to an additional bath (taken at zero-temperature for simplicity) at rate $\Gnr$, by a Liouvillian, $\Liou_{\Gnr}$, that we obtain from \eqref{eq:Liou_full} with the replacements $\Gamma\to\Gnr$ and $n_{\rm th}^{(m)}\to 0$.
Putting things together, the master equation reads:
\be
\dot{\rho}(t) =
-\frac{i}{\hbar}[H,\rho(t)]
+ \Liou_\Gamma \rho(t)
+ \Liou_{\deph} \rho(t)
+ \Liou_{\Gnr} \rho(t)
\ .
\ee
We obtain the steady-state density matrix by analytical solution of the algebraic equation $\dot{\rho}(t) = 0$.
From that solution and using input-output theory, the reflection coefficient can be written as
\be
r=1+\frac{2\Gamma}{\Omega} \sum_j \xi_j \rho_{j,j+1} \label{eq:rinout}
\ee

\subsection{Reflection coefficient in the small excitation limit}\label{app:Th:refl}

We take $N=3$, which is fully appropriate to describe scattering when $\Omega \lesssim \Gamma$.
We consider resonant scattering ($\delta=0$) and make a first-order expansion in the small quantities $\nge$, $\nef$, $\deph/\Gamma$, and $\Gnr/\Gamma$, and a second-order expansion in the ratio $\Omega/\Gamma$. We arrive at the following expression
\be
r(0)=-1 + \frac{8\nge+4\nef}{1+3i\Gamma/2|\alpha|} + 4 \frac{\deph}{\Gamma} + 2\frac{\Gnr}{\Gamma} + 
\frac{4(\Omega/\Gamma)^2}{1+i\Gamma/|\alpha|} \label{eq:r0full}
\ee
Taking  $\nth=\nge=\nef$ gives Eq.~\eqref{eq:r0} in the main text.

\subsection{Reflection coefficient for a two-level system}

Solving the master equation for a two-level system results in the following expression for the reflection coefficient
\be
r(0)_{\rm TLS}=-1+4
\frac{2 \nth+2 \nth^2+(\Omega/\Gamma)^2}{1+4\nth+4 \nth^2+ 2 (\Omega/\Gamma)^2} \ . \label{eq:rTLS}
\ee
Notice that a two-level system is not equivalent to a three-level system in the limit $\Grat \to 0$, because of the effect of thermal photons at the first-to-second excited state transition.
Equation \eqref{eq:rTLS} is plotted as a dotted line in Fig.~\ref{fig:noiseampsweep}(b).

\subsection{Sensitivity and responsivity}

Taking the derivative of the reflection coefficient at resonance in the limits $\Grat \to 0$ and $\nth \ll 1$ with respect to $\nth$ gives the following expression for the responsivity:
\be
\mathcal{R}=\frac{2 \left(6+\Omega _r^2\right)}{\left(1+2 \Omega _r^2\right){}^2} \ , \label{eq:Ranalyt}
\ee
where $\Omega_r=\Omega/\Gamma$.
By inserting \eqref{eq:Ranalyt} into \eqref{eq:netp}, we obtain the sensitivity
\be
\NETP= \frac{\left(1+2 \Omega _r^2\right){}^2}{\sqrt{2 \Gamma \eta } \Omega _r \left(6+\Omega _r^2\right)} \ .
\ee
These expressions are plotted as dashed lines in Fig.~\ref{fig:SensResp}.

\section{Data Analysis}\label{app:Analys}

\subsection{Noise windowing}

We introduce a weight function
\be
w(\omega)= \frac{8}{12} \Lorfun(\omega-\omega_{\rm ge},\Gamma) + \frac{4}{12}\Lorfun(\omega-\omega_{\rm ef},2\Gamma) \ ,
\ee
where $\Lorfun(\omega,\Gamma)$ is a normalized Lorentzian of linewidth $\Gamma$.
For a frequency-dependent noise profile $S(\omega)$, we calculate an effective added photon number as
\be
\delta\nth^{\rm eff} = \int d\omega w(\omega) S(\omega) .
\ee
For a 400 MHz white noise of amplitude $\delta \nth$ [Fig.~3], we find $\delta\nth^{\rm eff}= \delta\nth \wwin$, with $\wwin=0.89$.

\subsection{Conversion between thermal occupation and temperature}

If the waveguide is thermalized at a temperature $T_r$, the thermal occupation at a given frequency is given by the Bose distribution
\be
n(\omega,T_r)=\frac{1}{\exp(\hbar\omega/k_BT_r)-1}
\ee
The effective thermal occupation sensed by the thermometer as it appears in Eq.~\eqref{eq:r0}, taking into account Eq.~\eqref{eq:r0full}, is given by
\be
\nth(T_r)=\frac{2}{3} n(\omega_{\rm ge},T_r)+\frac{1}{3} n(\omega_{\rm ef},T_r) \label{eq:T2nth}
\ee
To convert thermal occupation into temperature, we numerically invert \eqref{eq:T2nth}, which is a monotonic function of $T_r$.

\subsection{Estimate of thermal occupation based on nominal attenuation}

To estimate the thermal occupation of the waveguide, we model each attenuator as a beam-splitter combining the incoming signal with a thermal state with occupation $n(\omega,T_{i,\rm{att}})$, corresponding to the temperature $T_{i,\rm{att}}$ of the stage at which the attenuator resides. The thermal occupation $n_i(\omega)$ after the $i$-th thermalization stage with attenuation $A_i$ is thus given by \cite{Krinner2019}
\be
n_i(\omega)=\frac{n_{i-1}(\omega)}{A_i}+\frac{A_{i}-1}{A_i} n(\omega,T_{i,\rm{att}}) \ .
\ee

\section{Influence of an imperfect circulator in the referenced reflection measurement}\label{app:circ}

As explained in the main text, we take advantage of the fact that we can ``switch off'' our emitter in-situ, either by detuning its frequency outside of the band of interest, or by saturating it with a strong coherent tone, to obtain a calibration trace for our measurement chain. Using this trace as a reference in our reflection measurements, we effectively cancel out the frequency-dependent transfer function of the chain. However, this procedure does not account for interference effects that can be introduced by imperfections in the circulator used to route the signals from the input line to the device under test (DUT), and from the DUT to the output line. In this section we model our measurement in the presence of an imperfect circulator, and provide an estimate for the systematic error that an imperfect circulator introduces in our temperature measurements.

Due to the finite isolation of the circulator, part of the probe signal will leak directly from the input port (port 1 in the following) to the output port (port 3), and interfere with the signal that has followed the intended path towards the DUT (port 2), then to the DUT and back, and finally to the output port.
We are interested in the relation between the reflection coefficient of the DUT, $\Gdut$, and the transmission coefficient, $\bar t$, measured across the network. Our configuration can be described as a three-port network with scattering parameters $s_{ij}$ (the circulator), with one port connected to a load with reflection coefficient $\Gamma_L$, and the other two ports matched. A simple analysis of the associated signal flow diagram \cite{Pozar2012} gives
\be
\bar t=AG\left(s_{31}+ \frac{s_{21} \Gamma_L s_{32}}{1-s_{22}\Gamma_L}\right) \ ,
\ee
where $A$ and $G$ stand for attenuation and gain before and after the circulator, respectively.
We write the scattering matrix of a lossless, symmetric, imperfect circulation with isolation $\gamma$ as
\be
s_{ij}=\begin{pmatrix}
-\gamma & \gamma & 1-\gamma^2 \\
1-\gamma^2 & -\gamma & \gamma \\
\gamma & 1-\gamma^2 & -\gamma
\end{pmatrix}_{ij}
\ee
In modeling the reflection coefficient of the load, we take into account electrical delay, $\tau$ and a phase shift $\phi$, so that:
\be
\Gamma_L(\omega)= \Gdut(\omega) \exp (i \omega \tau +i \phi)
\ee
We note in passing that the effect of cable attenuation between the circulator and the sample, which we neglect in the following, can be modeled by adding a small imaginary part to the phase $\phi$.
Finally, to model the referenced reflection coefficient, $r_{\rm refd}$, we take the ratio between the expected transmission when $\Gdut=r_{\rm ideal}$, $r_{\rm ideal}$ being the theory prediction, and $\Gdut=1$.
\be
r_{\rm refd} = \frac{\bar t(\Gdut=r_{\rm ideal})}{\bar t(\Gdut=1)} \ .
\ee

We now focus on the case of resonant reflection and small thermal occupation, for which $r_{\rm ideal} \approx -1+\delta_r$ and $\delta_r \approx 12 \nth \ll 1$. A series expansion in the small parameters $\gamma$ and $\delta_r$ thus gives for the real, imaginary part, and absolute value of $r_{\rm meas}$:
\begin{align}
\mbox{Re}[r_{\rm refd}] &= r_{\rm ideal}
+8 \gamma ^2 \sin^2 (\omega \tau +\phi ) +  2 \gamma  \delta _r \cos (\omega \tau  +\phi )
\\
\mbox{Im}[r_{\rm refd}] &=
-4 \gamma  \sin (\omega \tau  +\phi )+4 \gamma  \delta _r \sin (\omega \tau  +\phi ) \\
|r_{\rm refd}| &= |r_{\rm ideal}| -2 \gamma  \delta _r \cos (\omega \tau  +\phi )
\end{align}
Notice that the first correction is of first-order in the small quantities for the imaginary part, but only of second order for the real part and the absolute value.
By realistically choosing $\gamma=0.08$ (corresponding to 22dB isolation), and $\delta_r=0.06$ (or $\nth=5 \times 10^{-3}$), the maximum error on the real part due to interference effects is $0.05$ (or $4  \times 10^{-3}$ in thermal occupation), while the maximum error on the absolute value is $0.01$ (or $9 \times 10^{-4}$ in thermal occupation).


%

\end{document}